\documentstyle[12pt]{article} \newlength{\extraspace}
 \newlength{\extraspaces}
\begin{document} \pagestyle{empty} \begin{titlepage} \begin{flushright}
UTPT-94-20\\ hep-ph/9407311\end{flushright} \vspace{2.5cm} \begin{center}
{\LARGE Hints of new flavor physics at LEP?}\\\vspace{40pt} {\large B.
Holdom\footnote{holdom@utcc.utoronto.ca}} \vspace{0.5cm}

{\it Department of Physics\\ University of Toronto\\ Toronto,
Ontario\\Canada M5S 1A7} \vspace{0.5cm}

\vskip 2.1cm \vspace{25pt} {\bf ABSTRACT}

\vspace{12pt} \baselineskip=18pt \begin{minipage}{5in}

A massive gauge boson coupling to the third family but not lighter families
mixes with the $Z$.  The $Z$ partial width to $b\overline{b}$ increases,
the $\tau$ asymmetry parameter increases, the invisible width of the $Z$
decreases, and $\alpha_s(m_Z)$ decreases, all in a manner consistent with
present data.

\end{minipage} \end{center} \vfill \end{titlepage} \pagebreak
\baselineskip=18pt \pagestyle{plain} \setcounter{page}{1}

The large top mass may signal the presence of new gauge symmetries which
are broken close to the 1 TeV energy scale, and these gauge interactions
may also be felt by other members of the third family.  For example there
have been analyses of the effects of extended technicolor (ETC)
interactions on the $Zb\overline{b}$ vertex by relating these effects to
the physics responsible for the top mass.\cite{A}  This is the physics of
technicolored ETC gauge bosons which cause transitions between
technifermions and the third family.

In this note we would like to consider the effects of a gauge boson which
is a singlet under unbroken gauge symmetries and with a mass in the few
hundred GeV to one TeV range. Such a gauge boson, $X$, may correspond to a
diagonal generator of some broken flavor gauge symmetry.  The important
point is that it is typical for the last flavor symmetry to break to
involve the heaviest family, and thus the $X$ boson naturally couples to
the third family but not lighter families.  Its main effects\footnote{We
discuss fermion mass mixing effects at the end.} may arise through a small
mixing with the $Z$. These effects will differ from those of the more often
studied $Z'$ appearing in the grand unified context, since the $Z'$ couples
universally to all families of fermions.

An extra gauge boson can mix with the $Z$ through a mass mixing term and a
kinetic energy mixing term.  It can also mix with the photon through a
kinetic energy mixing.  These mixings show up in shifts in the standard
electroweak correction parameters, $S$, $T$, and $U$.  And since the mixing
induces small $X$ couplings to the light families proportional to the $Z$
and photon couplings, the standard neutral current processes are also
affected by $X$ exchange.  But all these effects \cite{B} are proportional
to the square of the mixing parameters.  On the other hand the physical
observables associated with $Z$ decay into third family fermions do receive
corrections to linear order in the mixing parameters.  It is appropriate to
consider these latter effects in light of the new data from $Z$ factories.

The vector and axial couplings of the $X$ boson to members of the third
family will be denoted by ${g}_{Xv}^{f}$ and ${g}_{Xa}^{f}$ with $f
=t,b,{\nu }_{\tau },\tau $.  Other fermions (but not light fermions) may
also carry $X$ charge, and when summed over all such fermions the $X$
charges are anomaly free and orthogonal to standard model hypercharges.
The standard model $Z$ couplings will be denoted by ${g}_{v}^{f}$ and
${g}_{a}^{f}$.

The mixing effects are described by the following Lagrangian quadratic in
the $A$, $Z$, and $X$ fields.  Primes are attached to indicate that these
fields, which have the couplings to matter just described, do not yet have
conventional kinetic or diagonal mass terms.  \begin{eqnarray} {{\cal
L}}_{AZX} & = & -{\frac{1}{4}}{A'}_{\mu \nu }{A'}^{\mu \nu
}-{\frac{1}{4}}{Z'}_{\mu \nu }{Z'}^{\mu \nu }-{\frac{1}{4}}{X'}_{\mu \nu
}{X'}^{\mu \nu }\nonumber\\& & +{\frac{1}{2}}{m}_{Z}^{2}{Z'}_{\mu
}{Z'}^{\mu }+{\frac{1}{2}}{m}_{X}^{2}{X'}_{\mu }{X'}^{\mu }\nonumber\\& &
+x{m}_{Z}^{2}{Z'}_{\mu }{X'}^{\mu }-y{\frac{1}{2}}{Z'}_{\mu \nu }{X'}^{\mu
\nu }-w{\frac{1}{2}}{A'}_{\mu \nu }{X'}^{\mu \nu } \label{b}\end{eqnarray}
We perform a transformation to unprimed fields which recovers conventional
kinetic terms and a diagonal mass matrix.  The full result to second order
in the mixing parameters $x$, $y$, and $w$ is given in \cite{B}.

The relation of interest here is \begin{equation} {X'}_{\mu }={X}_{\mu
}+(y-x){\frac{{m}_{Z}^{2}}{{m}_{X}^{2}}}{Z}_{\mu } \end{equation} where we
have kept only the first order terms in the mixing parameters and assumed
that ${m}_{X}^{2}\gg {m}_{Z}^{2}$.  When this substitution is made in the
full Lagrangian we find that the $Z$ couplings to the third family are
shifted by amounts \begin{equation}\delta
{g}_{v,a}^{f}=(y-x){\frac{{m}_{Z}^{2}}{{m}_{X}^{2}}}{g}_{Xv,a}^{f}.\label{c}
\end{equation}

The best constraints on these couplings come from the partial decay widths,
${\Gamma }_{f}$, and the asymmetry parameters,
\begin{equation}{A}_{f}={\frac{2{g}_{v}^{f2}{g}_{a}^{f2}}{{g}_{v}^{f2}+{g}_{
a}^{f2}}}.\end{equation}
They are shifted by amounts \begin{equation}{\frac{\delta {\Gamma
}_{f}}{{\Gamma }_{f}}}={\frac{2({g}_{v}^{f}\delta
{g}_{v}^{f}+{g}_{a}^{f}\delta
{g}_{a}^{f})}{{g}_{v}^{f2}+{g}_{a}^{f2}}},\label{k}\end{equation}
\begin{equation}{\frac{{\delta
A}_{f}}{{A}_{f}}}={\frac{{{g}_{a}^{f}}^{2}-{{g}_{v}^{f}}^{2}}{{{g}_{a}^{f}}^
{2}+{{g}_{v}^{f}}^{2}}}
\left({{\frac{{\delta g}_{v}^{f}}{{g}_{v}^{f}}}-{\frac{{\delta
g}_{a}^{f}}{{g}_{a}^{f}}}}\right).\label{l}\end{equation}  The vector $Z$
coupling to $\tau$ is small, ${g}_{v}^{\tau }\approx 0.07{g}_{a}^{\tau }$,
and thus ${\Gamma }_{\tau }$ is really only sensitive to a shift in
${g}_{a}^{\tau }$. ${\Gamma }_{\tau }$ has been found to be consistent with
lepton universality to a fraction of a percent, thus making ${g}_{a}^{\tau
}$ the most constrained of the third family $Z$ couplings. If we wish this
not to constrain the $Z$-$X$ mixing then we must ensure that the $X$ boson
has only vector couplings to the $\tau$.

To proceed further we will consider an explicit model in which the $X$
charges of the third family are the following.
\begin{equation}\begin{array}{c}{g}_{Xv}^{t,b} =
0\;\;\;\;\;,\;\;{g}_{Xa}^{t,b} = g_X\\ {g}_{Xv}^{\tau } =
g_X\;\;\;,\;\;\;{g}_{Xa}^{\tau } = 0\;\\{g}_{XL}^{\nu_\tau} \equiv
{g}_{Xv}^{\nu_\tau}+{g}_{Xa}^{\nu_\tau}=
g_X\end{array}\label{a}\end{equation} The model we have in mind is a
recently proposed dynamical model for the large top mass, in which the
quarks are required to have axial couplings to the $X$ boson.\cite{C} We
will find that the source of $Z$-$X$ mixing in the model also relies on the
axial $X$ couplings to quarks.

We first present the results.  In Table 1 we compare the measured shifts of
five observables to the expected shifts due to the $X$ boson.  The relative
sizes and relative signs of the expected shifts are completely determined,
given the $X$ charges in (\ref{a}).  The model then provides the additional
input to determine the overall sign and to allow an estimate of the overall
magnitude of the shifts.

\begin{table}\centering\large $\begin{array}{|c|l|l|}\hline & {\rm
Measurement} & X\;{\rm boson} \\ \hline{{{\delta\Gamma }_{b}}/{{\Gamma
}_{b}}} & \!\!\!\begin{array}{l}a)\;+0.031\pm 0.011\\ \;\;\;\;\;+\,0.030\pm
0.014\end{array} & +\,0.021 \\ \hline{{{\delta\Gamma }_{{\nu }_{\tau
}}}/{{\Gamma }_{{\nu }_{\tau }}}} & b)\;-0.014\pm 0.023 & -\,0.015\\ \hline
{{{\delta\Gamma }_{\tau }}/{{\Gamma }_{\tau }}} & c)\;+0.002\pm 0.005 &
+\,0.0022 \\ \hline {{{\delta A}_{\tau }}/{{A}_{\tau }}} & d)\;+0.31\pm
0.13 & +\,0.21 \\ \hline {{{\delta A}_{b}}/{{A}_{b}}} & e)\;-0.02\pm 0.16 &
-\,0.0054 \\ \hline \end{array}$\caption{These shifts are due to
universality breaking effects only, and do not include effects already
contained in the standard model.}\end{table}

In the measurements we isolate those departures from universality beyond
those already contained in the standard model.  We are not interested in
univerality preserving corrections such as oblique corrections from new
physics.  The measured shifts in Table 1 do not include oblique corrections
because we are careful to extract the shifts from observables insensitive
to oblique corrections.   In particular, the values in the table labeled
$a$) to $e$) were obtained as follows.

\noindent $a$) We use ${\Gamma }_{b}/{\Gamma }_{\rm had}={\Gamma
}_{b}/({\Gamma }_{b}+{\Gamma }_{{\rm had}\neq b})$ and compare its value,
from \cite{D} and \cite{F} respectively, to the standard model value of
0.2158 for a top mass of 175 GeV.

\noindent $b$) We extract the value of ${\Gamma }_{\rm inv}/{\Gamma }_{\ell
}=({\Gamma }_{{\nu }_{e}}+{\Gamma }_{{\nu }_{\mu }}+{\Gamma }_{{\nu }_{\tau
}})/{\Gamma }_{\ell }$ from \cite{D} using the near $e$-$\mu$-$\tau$
universality for ${\Gamma }_{\ell }$, and compare to a standard model value
of 5.976.

\noindent $c$) We extract the value of ${\Gamma }_{\tau }/{\Gamma }_{e,\mu
}$ from \cite{D} assuming $e$-$\mu$ universality and compare to unity.

\noindent $d$) The ratios ${A}_{FB}^{0\tau }/{A}_{FB}^{0e,\mu }$ from
forward-backward measurements and ${3P}_{\tau }/{4P}_{\tau }^{FB}$ from
$\tau$ polarization measurements are both equal to ${A}_{\tau }/{A}_{e}$
assuming $e$-$\mu$ universality.  We obtain these two ratios from  \cite{F}
and then average them and compare to unity.  This departure from unity
appears to be as significant as the more publicized ${{\delta\Gamma
}_{b}}/{{\Gamma }_{b}}$.  (Caution:  SLD reports a value for $A_e$
inconsistent with and larger than the LEP values.)

\noindent $e$) We use the value of ${A}_{FB}^{0b}/{A}_{FB}^{0c}$ from
\cite{D,F} and compare to a standard model fit value of 1.4  \cite{F}.  The
errors, especially in ${A}_{FB}^{0c}$, make this uninteresting.

The $X$ boson also leads to a decrease in the value of ${\alpha
}_{s}({m}_{Z})$ extracted from the measurement of ${R}_{\ell }={\Gamma
}_{\rm had}/{\Gamma }_{\ell }$.   This is because ${R}_{\ell }$ remains the
same if the increase in ${\Gamma }_{b}$ is compensated by a decrease in the
QCD corrections to ${\Gamma }_{\rm had}$.   Using the ${\alpha }_{s}$
dependence of ${R}_{\ell }$ in \cite{H} we find that the shift is $\delta
{\alpha }_{s}({m}_{Z})=-0.014$. Shifting the value of ${\alpha
}_{s}({m}_{Z})$ in \cite{D,F} by this amount gives \begin{equation}{\alpha
}_{s}({m}_{Z})=0.110\pm 0.007.\end{equation} This is typical of values of
${\alpha }_{s}({m}_{Z})$ obtained from deep inelastic scattering and heavy
quarkonia decays,  and is lower than values from jet studies.\cite{I}

We now briefly describe the model \cite{C} and derive our results.  The
model contains a fourth family, and members of the third and fourth
families are composed from two ``families" of fermions $f$ and
$\underline{f}$.  The massive $X$ boson, which appears when a nonabelian
hypercolor gauge group breaks, couples with a vector charge of $+g_X$ to
all members of the $f$ family and with a vector charge of $-g_X$ to all
member of the $\underline{f}$ family.  Each of these families has standard
$SU(3)\times SU(2)\times U(1)$ quantum numbers.  But these fields are not
the mass eigenstates.  The mechanism producing a large top mass requires
that the fourth family quark mass eigenstates $t'$ and $b'$ correspond to
Dirac spinors of the form $[{\underline{f}}_{L},{f}_{R}]$.  The $t'$ and
$b'$ are nearly degenerate with masses close to a TeV.  The $t$ and $b$
quarks correspond to $[{f}_{L},{\underline{f}}_{R}]$, which then implies
that the $X$ boson couples with the same axial coupling to the $t$ and $b$
quarks.

We will see that this leads to a positive shift in ${\Gamma }_{b}$. This
prediction is to be compared with the negative shift found in the simplest
extended technicolor theories.\cite{A}  The analog of this negative ETC
contribution does not exist in our model since there is no gauge boson
which causes a transition between a $b$ and a more massive fermion other
than the $t$. At one loop, the only nonstandard correction to the
$Zb\overline{b}$ vertex is due to $Z$-$X$ mixing.

The mechanism by which the top receives a large mass need not also be
occurring in the lepton sector, and thus the lepton mass eigenstates were
not determined in \cite{C}.  What is happening in the lepton sector is
uncertain, and thus we shall simply assume that the dynamics of the model
is such that the leptonic $\underline{f}$ fields describe the fourth family
leptons (with masses somewhat less than the $t'$ and $b'$ masses).  The
$\tau $ and the ${\nu }_{\tau }$ are then described by leptonic $f$ fields;
this implies for example the vector $X$ coupling to the $\tau$.  Notice
that the leptonic $X$ charges would change sign if the dynamics of the
model led to the reverse identification of the $\underline{f}$ and $f$
fields.  This would change the overall sign of the shifts in the leptonic
observables.

The universality breaking effects are ultimately caused by $X$ boson
exchange, and thus we need the value of $g_X^2/m_X^2$.  The model provides
an estimate since the main source of mass for both the $X$ and $Z$ bosons
is the dynamical $t'$ and $b'$ masses.  The ratio of the $X$ and $Z$ masses
is determined by the ratio of their respective axial couplings to these
quarks. \begin{equation}{\frac{{g}_{X}^{2}}{{m}_{X}^{2}}}
={\frac{{\left({{\frac{e}{4cs}}}\right)}^{2}}{{m}_{Z}^{2}}}\label{d}
\end{equation}
The $Z$ and to a lesser extent the $X$ also receive a small contribution to
their mass from the fourth family leptons.  This would reduce
${m}_{X}/{m}_{Z}$ and thus slightly increase our estimate of
${g}_{X}^2/{m}_{X}^2$.  $g_X$ must be large enough to ensure that
${m}_{X}^{2}/{m}_{Z}^{2}\gg 1$, as we have previously assumed.  This is
reasonable since the $X$ boson emerges from a strongly interacting
hypercolor interaction.

Because of the $X$ couplings a quark loop can produce $Z$-$X$ mass mixing
only if there is isospin violating quark masses.  The main motivation for
the model was to show how $t'$ and $b'$ can remain almost degenerate and
thus produce little contribution to $\Delta\rho$.  Similarly, the $t'$ and
$b'$ contributions to $Z$-$X$ mass mixing will largely cancel.  The $\tau
'$ loop contribution nearly vanishes since the $Z$ and $X$ have essentially
axial and vector $\tau '$ couplings respectively.  The massive ${\nu
}_{\tau '}$ could contribute, but its mass is also constrained by
$\Delta\rho$.  Thus it is fair to assume that the $t$ loop is the main
source of $Z$-$X$ mass mixing.

We may determine the mixing parameter $x$ as defined in (\ref{b}) by taking
the ratio of the $Z$-$X$ mixing diagram (a $t$ loop) to the $Z$ mass
diagram (a $q'$ loop).  These loop diagrams are convergent and are
dependent on the momentum-dependent dynamical mass functions for the $t$
and $q'$. Neither of these masses arise from a simple ETC boson exchange,
and thus both mass functions are expected to fall with increasing momentum
more quickly than quark masses in conventional ETC theories.  If one mass
function was just the scaled version of the other mass function then the
result for $x$ would be \begin{equation} x ={\frac{{\frac{e}{4cs}}{g}_{X}
{{m}_{t}}^{2}}{{\left({{\frac{e}{4cs}}}\right)}^{2}  2
{{m}_{q'}}^{2}}}.\label{e}\end{equation}  The factor of 2 is for the two
flavors $t'$ and $b'$.  We will absorb the uncertainty in the mass
functions into the uncertainty in the value of $m_{q'}$.  In \cite{C} we
estimated $m_{q'}\approx 1$ TeV.

The mixing parameter $y$ receives contributions from a $b$ loop and a ${\nu
}_{\tau }$ loop.  These loops are log divergent and they are cutoff by the
$t$ mass and ${\nu }_{\tau '}$ mass respectively.  By setting a logarithm
equal to unity we estimate \begin{equation} y\approx{\frac{e}{4 c s}}
{\frac{{g}_{X}}{6 {\pi }^{2}}}.\end{equation}  This turns out to be less
than 4\% of $x$, and thus we will simply ignore $y$.

By combining the results in (\ref{c}), (\ref{a}), (\ref{d}), and (\ref{e})
we find \begin{equation}\begin{array}{c}{\delta g}_{v}^{t,b} =
0\;\;\;\;\;,\;\;{\delta g}_{a}^{t,b} = Y\\ {\delta g}_{v}^{\tau } =
Y\;\;\;,\;\;\;{\delta g}_{a}^{\tau } = 0\;\\{\delta g}_{L}^{\nu } =
Y\end{array}\label{p}\end{equation} with \begin{equation}
Y=-{\frac{e}{8cs}}
{\left({{\frac{{m}_{t}}{{m}_{q'}}}}\right)}^{2}.\end{equation} Notice that
we have not needed to determine $m_X$ or $g_X$ separately; and although the
sign of $x$ depends on the sign of $g_X$, the final result does not.  By
inserting (\ref{p}) along with ${m}_{t}=175$ GeV, ${m}_{q'}=1$ TeV, and
$s^2=.232$ into (\ref{k}) and (\ref{l}) gives the shifts in Table 1. Also,
by combining our results with those of \cite{B} we have confirmed that the
$Z$-$X$ mixing has a negligible effect on oblique parameters $S$, $T$, and
$U$ and neutral current processes.  Among these, the largest change is a
$0.03$ increase in $T$.

Finally, an $X$ boson having nonuniversal flavor couplings is an obvious
source of flavor changing neutral currents (FCNCs) among light quarks,
because of the fermion mass mixing which must occur between families.  For
example an $X$ boson exchange produces an operator ${\overline{t}}{\gamma
}_{\mu }\gamma_5{t}{\overline {t}}{\gamma }^{\mu }\gamma_5{t}$.  When
expressed in terms of the quark mass eigenstates there may be a
contribution to the $\Delta C=2$ operator ${\overline{c}}{\gamma }_{\mu
}\gamma_5{u}{\overline {c}}{\gamma }^{\mu }\gamma_5{u}$.  The size of this
flavor changing effect depends on how the mass mixing arises in the model.
If, as suggested in \cite{C}, all mass mixing between families arises in
the up sector then the resulting $D^0-\overline{D}^0$ mixing will involve a
factor of order ${\left|{V_{ub}V_{cb}}\right|}^{2}$ where $V$ is the KM
matrix.  This provides more than adequate suppression.  Mass mixing would
only be transmitted to the down sector via additional weak effects, and
this would keep the contributions to $K^0-\overline{K}^0$ and
$B^0-\overline{B}^0$ mixing at a safe level.  This type of scenario has
been described \cite{J} in the context of the usual FCNC problem of
technicolor theories.

The $X$ boson we have described causes a distinctive pattern of
universality breaking corrections.  The comparison with the data in Table 1
is tantalizing, and we hope that this motivates further interest in the
search for new flavor physics in precision experiments.

\newpage \vspace{3ex} \noindent {\Large\bf Acknowledgments} \vspace{1ex}

I thank D. Bailey for discussions.  This research was supported in part by
the Natural Sciences and Engineering Research Council of Canada.

\end{document}